\documentclass[seceq]{ptptex}
\usepackage{wrapft}

\usepackage{graphicx}

\def\fsl#1{\setbox0=\hbox{$#1$}           % set a box for #1 
   \dimen0=\wd0                                 % and get its size
   \setbox1=\hbox{/} \dimen1=\wd1               % get size of /
   \ifdim\dimen0>\dimen1                        % #1 is bigger
      \rlap{\hbox to \dimen0{\hfil/\hfil}}      % so center / in box
      #1                                        % and print #1
   \else                                        % / is bigger
      \rlap{\hbox to \dimen1{\hfil$#1$\hfil}}   % so center #1
      /                                         % and print /
   \fi}          
\title{%        %You can use \\ for explicit line-break
Quest for the Dynamical Origin of Mass
\footnote{Talk presented on Jan. 26, 2009 at 
    Yukawa International Seminar (YKIS)
 "Particle Physics beyond the Standard Model"
 Yukawa Institute for Theoretical Physics (YITP),Kyoto University, Japan, January 26 - March 25, 2009,
}
\subtitle{An LHC perspective from Sakata, Nambu and Maskawa}    %Use this when you want a subtitle.

\author{%       %Use \scshape  for the family name
Koichi \textsc{Yamawaki}%$^{1,}$
}

\inst{%         %Affiliation, neglected when [addenda] or [errata]
Department of Physics, Nagoya University, Nagoya 464-8602, Japan
}

}

\abst{%         %this abstract is neglected when [addenda] or [errata]
I review the dynamical symmetry breaking (DSB) approach to the Origin of Mass, which is traced back to the original  (2008 Nobel prize) work of Nambu based on the BCS analogue of superconductor  where mass of nucleon (then elementary particle) arises due to Cooper paring  and pions are provided as massless Nambu-Goldstone (NG) bosons, being composite as in Fermi-Yang/Sakata model.  
In this talk I will focus on the modern version of DSB or composite Higgs models: Walking/Conformal Technicolor,  Hidden Local Symmetry (HLS) or Moose,  and Top Quark Condensate, with the their extra dimension versions closely related with HLS.  Particular emphasis will be placed on the large anomalous dimension and conformal symmetry at the conformal fixed points, developed along the line of the pioneering work of Maskawa and Nakajima. Due to (approximate) conformal symmetry these models  do have composite Higgs particle (``Techni-dilaton'' , ``Top-sigma'' etc.).  Weakly coupled composite gauge boson is realized at ``Vector Manifestation'' formulated at conformal fixed point, which may be applied to the composite W/Z boson models. They will be tested 
in the upcoming LHC experiments.  

}

\begin{document}

\maketitle

\section{Introduction}
The most urgent problem of the modern particle theory is to reveal the Origin of Mass. 
 In the Standard Model (SM)  the spontaneous symmetry breaking (SSB) of the electroweak symmetry  is attributed to a single parameter, $v =246\, {\rm GeV}$,
 the vacuum expectation value (VEV) of the field of a hypothetical elementary particle, the Higgs boson, which then is distributed via
 gauge and Yukawa couplings $g_i$ to all the  particles having mass $m_i$: $m_i \sim g_i v$. Yet the nature of Higgs boson remains mysterious. In order to develop the VEV the mass squared of the Higgs in the Lagrangian  should be tuned 
 {\it negative (tachyon!)}  on the weak scale  in an ad hoc manner. 
Particle theorists looking desperately beyond the SM have been fighting  on this central problem over 30 years without decisive experimental
information.  Now we are facing a new era that LHC experiments  will tell us which
theory is right while others are not.

It should be recalled that the very concept of SSB was created by the 2008 Nobel prize work of Nambu~\cite{Nambu60,NJL} in the concrete form of  DSB where the nucleon mass was dynamically generated via Cooper pairing of (then elementary) nucleon and anti-nucleon, ``nucleon condensate'',  based on the Bardeen-Cooper-Schrieffer (BCS) analogue of superconductor: Accordingly, there appeared pions as massless Nambu-Goldstone (NG) bosons which were dynamically generated to be nucleon composites in the same sense as in the Fermi-Yang/Sakata model~\cite{FY}. 
{\it Thus the SSB was born as DSB!} Before advent of the concept of SSB, low energy hadron physics was well described by 
the effective theory of  Gell-Mann-Levy (GL) linear sigma model 
with an elusive scalar boson, the sigma meson, 
which was  unjustifiably assumed to have negative mass squared. 
Actually,  the GL linear sigma model Lagrangian is  a model equivalent to the SM Higgs Lagrangian.

The real physical meaning of this mysterious tachyonic mode was actually revealed by Nambu
 as the BCS instability where attractive forces between nucleon and anti-nucleon give rise to the
nucleon Cooper paring (tachyonic bound state) which changes the vacuum from the original (free) one into  the true one
having no manifest symmetry. 
The Nambu's theory for the nucleon mass was later developed into DSB in the underlying microscopic theory QCD where the gluonic attractive forces again generate the Cooper paring of quark and antiquark (instead of nucleon and anti-nucleon), quark condensate $\langle \bar q q \rangle$, which then gives rise to  the BCS instability and the dynamical mass of quarks: Pions are now quark composites. Hence Nambu's idea was established in a deeper level of matter.

This DSB in QCD is the prototype of the Technicolor (TC)~\cite{TC} .
Just in the same way as the GL sigma model was for QCD, the SM Higgs Lagrangian  may be regarded as an effective theory for the  hypothetical underlying gauge theory like QCD, ``Technicolor'': Higgs boson may be regarded as the composite particle like the sigma meson (``techni-sigma''). In much the same way as the sigma meson condensate $\langle \sigma\rangle =f_\pi  (=93 {\rm MeV}) \sim \Lambda_{\rm QCD}$ was an effective description of the quark condensate $\langle \bar q q \rangle \sim \Lambda_{\rm QCD}^3$ in QCD, the Higgs condensate in SM $\langle H \rangle =F_\pi (=246 {\rm GeV}) \sim \Lambda_{\rm TC}$ would be replaced by the ``techni-fermion'' condensate $\langle \bar F F \rangle \sim \Lambda_{\rm TC}^3$ in TC, where $\Lambda_{\rm QCD}$ and 
$\Lambda_{\rm TC}$ are intrinsic scales of  the respective theories, with roughly a scale up of $\Lambda_{\rm TC}
\sim (F_\pi/f_\pi)\Lambda_{\rm QCD} \sim 2600 \cdot (250 {\rm MeV})\sim 700 {\rm GeV}$.

In order to accommodate mass of quarks/leptons $m_{q/l}$, we should further introduce interactions between the technifermion and the quarks/leptons. This is most typically done by Extended Technicolor (ETC)~\cite{Dimopoulos:1979es},~\footnote{
The same can be done in a composite model where quarks/leptons and technifermions are composites on the same footing.\cite{Yamawaki:1982tg}
} which yields mass of quarks/leptons $m_{q/l} \sim \frac{1}{\Lambda_{\rm ETC}^2}\, \langle \bar F F \rangle_{\Lambda_{\rm ETC}}$, where $\langle \bar F F \rangle_{\Lambda_{\rm ETC}} $ is the condensate evaluated at the scale of ETC $\Lambda_{\rm ETC} (\gg \Lambda_{\rm TC})$,  which would be  $\langle \bar F F \rangle_{\Lambda_{\rm ETC}} \sim \langle \bar F F \rangle_{\Lambda_{\rm TC}} \sim \Lambda_{\rm TC}^3 $ if  the TC is a simple scale-up of  QCD.
Then we would have  $m_{q/l} \sim \Lambda_{\rm TC}^3/\Lambda_{\rm ETC}^2 < (700)^3/ (10^6)^2 {\rm MeV} \sim 
0.3 {\rm MeV}$, if  we impose a constraint $\Lambda_{\rm ETC} > 10^{6} {\rm GeV}$ in order to avoid the  excessive
 Flavor-Changing-Neutral-Currents (FCNC).  Then the typical mass (s-quark mass) would be roughly $10^{-3}$ smaller than the reality.  
To avoid this problem, Holdom\cite{Holdom:1981rm} 
simply assumed that the  TC
has an
ultraviolet fixed point and the anomalous dimension becomes % near the fixed point, and iff the anomalous dimension is 
larger than unity $\gamma_m>1$ in the ultraviolet region so that the technifermion condensate at ETC scale is
enhanced $\langle \bar F F \rangle_{\Lambda_{\rm ETC}} = Z_m^{-1} \langle \bar F F \rangle_{\Lambda_{\rm TC}} $
where $Z_m^{-1} = \left(\Lambda_{\rm ETC}/\Lambda_{\rm TC}\right)^{\gamma_m} $, more than  $10^3$ times the simple scale up of  QCD which has a
vanishingly small anomalous dimension $\gamma_m\simeq 0$.

\begin{wrapfigure}{r}{6.6cm}
\vspace{-0.5cm}
  \includegraphics[width=6cm]{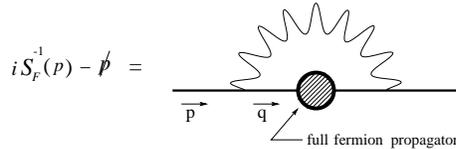}
\caption{Graphical expression of the SD equation 
         in the ladder approximation.}
\label{fig:SDeq}
\end{wrapfigure}

 The Holdom' mechanism unfortunately had no support by concrete dynamical arguments and no prediction for the value of the anomalous dimension.
It is my paper with M. Bando and K. Matumoto \cite{Yamawaki:1985zg} (receipt date on December 24,1985) 
that did demonstrate existence of such a theory having a concrete value of the large anomalous dimension $\gamma_m =1$ as desired, based  on the Spontaneous Chiral Symmetry Breaking (S$\chi$SB) solution of the ladder Schwinger-Dyson (SD) equation for fermion full propagator $S_F(p)$ parameterized as
 $i S_{F}^{-1} = A(p^2) \fsl{p} - B(p^2)$  with {\it non-running}  (ideal limit
of the ``walking'')  gauge coupling, $\alpha(Q) \simeq \alpha = {\rm constant}$. (See Fig. \ref{fig:SDeq}) 
 In 1986 similar enhancement effects of the condensate were also studied~\cite{Akiba} within the same ladder SD equation, without use of the renormalization-group equation (RGE) concepts of anomalous dimension and fixed point, rather emphasizing 
the asymptotic freedom of the TC theories with walking coupling.

 The ladder SD equation with non-running coupling was first consistently analyzed, with an explicit cutoff,  by Maskawa and Nakajima~\cite{Maskawa:1974vs} : 
They discovered that the SSB can only take place for strong coupling $\alpha>\alpha_{\rm cr} 
={\cal O} (1) $,  {\it non-zero critical coupling}.\footnote{
Earlier works\cite{JBW} in the ladder SD equation with non-running coupling all confused explicit
breaking solution with the SSB solution and thus implied $\alpha_{\rm cr}=0$. 
}  
From  the explicit form~\cite{Fukuda:1976zb}  of the Maskawa-Nakajima S$\chi$SB solution of the fermion mass function
$\Sigma(Q) =B(p^2)/A(p^2)$ in Landau gauge ($A(p^2)\equiv 1$), we found~\cite{Yamawaki:1985zg} :  
\begin{eqnarray}
\Sigma (Q) \sim 1/Q \quad (Q\equiv \sqrt{-p^2}) \gg \Lambda_{\rm TC}) \quad {\rm at} \,\,\, \alpha \rightarrow \alpha_{\rm cr},\, %,\\
\label{asymp}
\end{eqnarray}
and 
\begin{eqnarray}
\gamma_m &=& 1 \, \nonumber ,\\
m_{q/l} &\sim& \Lambda_{\rm TC}^2/\Lambda_{\rm ETC}\, ,
\label{gammam}
\end{eqnarray}
in comparison  with the operator product expansion $\Sigma (Q) \sim 1/Q^2\cdot (Q/\Lambda_{\rm TC})^{\gamma_m}$
and $\langle \bar F F \rangle_{\Lambda_{\rm ETC}}= - {\rm Tr} S_F(p)  \sim  \Lambda_{\rm ETC} \Lambda_{\rm TC}^2$
or $Z_m^{-1} =(\Lambda_{\rm ETC}/ \Lambda_{\rm TC})^1$,
 where  the critical coupling $\alpha_{\rm cr}$ was identified with a nontrivial UV stable fixed point of the  RGE  a la Miransky \cite{Miransky:1984ef};  $\alpha=\alpha(\Lambda)  \rightarrow \alpha_{\rm cr}$
as $\Lambda \rightarrow \infty$,  to keep finite the solution $\Sigma(0)$\cite{Fukuda:1976zb, Miransky:1984ef} 
 \begin{equation}
%\Sigma (0) 
\Sigma(0)
\sim \Lambda \, \exp \left(- \pi/\sqrt{\alpha/\alpha_{\rm cr} -1}
\right)   \, ,
\label{Miransky}
\end{equation}
which is often called ``Miransky scaling'' with an {\it essential singularity} at $\alpha=\alpha_{\rm cr}$ .
Since the ladder SD equation is {\it scale-invariant} (except for the explicit cutoff), with the critical coupling  
identified as the conformal fixed point, 
  we called the theory ``{\it Scale-invariant Technicolor}'' (would be  ``Conformal
  TC'' in a currently fashionable language) and  predicted a ``{\it Techni-dilaton}'', a relatively light Higgs-like composite object  due to approximate conformal symmetry.

Today the ``Walking/Conformal TC'' is simply characterized by near conformal property 
with  $\gamma_m \simeq 1$ (For a review see
Ref. \cite{Hill:2002ap}).  
Such a theory should have an almost non-running  and strong gauge coupling  (larger than a certain non-zero critical coupling for S$\chi$SB)  to be realized
either at UV fixed point  
or  IR fixed point, or  both (``fusion''  of  the IR and UV fixed points),  as was characterized by 
``{\it Conformal Phase Transition (CPT)}''~\cite{Miransky:1996pd}. In contrast to the simple QCD scale-up which is widely believed to have no composite  Higgs particle (``higgsless''),  a salient feature of the walking/conformal TC is the 
prediction of the composite Higgs as a techni-dilaton~\cite{Yamawaki:1985zg}.

In this talk I shall describe the DSB with  Large Anomalous Dimension (see Ref. \cite{Yamawaki:1996vr} for basis and classics before 1996), namely a class of composite models based on the walking/conformal gauge theories having large anomalous dimension characteristic to the conformal UV/IR fixed point.\\

\begin{itemize}
\item Walking/Conformal TC \\
Modern version \cite{Appelquist:1996dq, Miransky:1996pd} realized through the Banks-Zaks (BZ) IR fixed point
 \cite{Banks:nn}, $\alpha(Q) \simeq \alpha_*$, of the ``{\it Large $N_f$ QCD}'' .
 
Several issues:\\
$\cdot$  Phase Structure, or CPT~\cite{Miransky:1996pd} \\
$\cdot$ Top Quark Mass in the Walking/Conformal TC with $\gamma_m >1$\cite{Miransky:1988gk}\\
$\cdot$  Techni-dilaton and Light Composite Spectra\cite{Harada:2003dc}\\
$\cdot$  $S$ Parameter in the Walking/Conformal TC \cite{Harada:2005ru} \\

\item Hidden Local Symmetry (HLS)\cite{Bando:1984ej,Harada:2000kb}  and Holography in the  Walking/Conformal Theories \\
$\cdot$  ``Vector Manifestation'' \cite{Harada:2000kb}  at CPT \\
--Realization of  a  Weakly Coupled Composite Gauge Boson--\\
$\cdot$  Holographic Walking/Conformal TC\cite{Haba:2008nz} as an Extension of HLS\\
\item Top Quark Condensate (Top Mode Standard Model, TMSM) \cite{Miransky:1988xi,Nambu89,BHL:1989ds} \\
$\cdot$ TMSM in Higher Dimensions at UV fixed point\cite{HTY:2000uk}.\\
$\cdot$ Top Mode Walking/Conformal TC\cite{Fukano:2008iv}\\

\end{itemize}
All these models are based on the chiral phase transition (dynamical symmetry breaking) near the conformal region (``conformal window'') associated with
the UV/IR fixed point, which inevitably develops large anomalous dimension due to strong coupling.
I expect that they will be tested in the upcoming  LHC experiments.

\section{Walking/Conformal Technicolor}
\subsection{Large $N_f$ QCD as a walking/conformal TC}
Modern version  \cite{Appelquist:1996dq,Miransky:1996pd} of the walking/conformal TC is based on the BZ IR fixed point \cite{Banks:nn}  in the ``large $N_f$ QCD'' 
which is the QCD with many  flavors $N_f  \, (\gg 3)$ of  massless ``quarks'' (fundamental color representation)\footnote{
For walking/conformal TC based on higher representation/other gauge groups see, e.g.,  Ref. \cite{Sannino:2004qp} 
} : The  two-loop beta function is %given by
  $\mu \frac{d}{d \mu} \alpha(\mu) 
  = -b \alpha^2(\mu) - c \alpha^3(\mu) ,
$
where
$  b = \left( 11 N_c - 2 N_f \right)/(6 \pi)$, 
 $ c =\left[ 34 N_c^2 - 10 N_f  N_c 
      - 3 N_f  (N_c^2 - 1)/N_c \right] /(24 \pi^2)$ .
 When $b>0$ and $c<0$, i.e.,  $  N_f^* < N_f < \frac{11}{2} N_c $ 
($N_f^\ast \simeq 8.05$ for $N_c = 3$), there exists an IR fixed point (BZ  IR fixed point)
  at $\alpha=\alpha_*$ where
\begin{equation}
  \alpha_\ast =\alpha_*(N_c, N_f)  = - b/c .
\label{eq:alpha_IR}
\end{equation}
Note that $\alpha_* =\alpha_*(N_c, N_f) \rightarrow 0$ as $N_f \rightarrow  11 N_c/2$ and hence there exists a certain range $N_f^{\rm cr} <N_f < 11 N_c/2$  (``Conformal Window'')  satisfying $\alpha_* < \alpha_{\rm cr}$, where the gauge coupling $\alpha(Q) \, (< \alpha_*)$ gets so weak
that attractive forces are no longer strong enough to trigger the S$\chi$SB, namely the chiral symmetry gets restored and de-confinement takes place (non-Abelian Coulomb phase).
Here 
$\alpha_{\rm cr}$ may be evaluated by the ladder SD equation~\cite{Fukuda:1976zb}, $\alpha_{\rm cr} =  \pi/(3C_2)
=(\pi/3) (2 N_c/(N_c^2-1)) $, in which case $N_f^{\rm cr}$ is evaluated by the condition
$\alpha_*(N_c, N_f) = \alpha_{\rm cr}$, yielding 
 $N_f^{\rm cr} \simeq 4 N_c$ ($ =12$ for $N_c=3$) \cite{Appelquist:1996dq}. 
\footnote{
The value should not be taken seriously, since $\alpha_\ast=\alpha_{\rm cr}$ is of  $\cal {O} $(1) and
the perturbative estimate of  $\alpha_*$  is not so reliable there, although the chiral symmetry restoration in large $N_f$ QCD has been supported by many other arguments, most notably the lattice QCD simulations, which however
 suggest diverse results as to $N_f^{\rm cr}$; Some recent results do
$8<N_f^{\rm cr}<12$ (Kogut-Susskind  fermion),\cite{Appelquist:2009ty}
while other does $6<N_f^{\rm cr}<7$ (Wilson fermion) ~\cite{lattice}.
}
Thanks to the IR fixed point the gauge coupling is actually walking $\alpha(Q) \simeq \alpha_*$ over the range 
$0< Q< \Lambda$  for the conformal window, where $\Lambda$ is a two-loop RG invariant 
analogue of the $\Lambda_{\rm QCD}$.

Here we are interested in the SSB phase slightly off the conformal window, 
$0< \alpha_\ast - \alpha_{\rm cr}\ll 1$ ($N_f \simeq N_f^{\rm cr}$)  
where the fermion gets a tiny mass $\Sigma (0)$ (or more properly, $m$ such that $\Sigma(m)=m$)
 from the S$\chi$SB in
the form :\cite{Appelquist:1996dq}
\begin{equation}
%\Sigma (0) 
m 
\sim \Lambda \, \exp \left(- \pi/\sqrt{\alpha_*/\alpha_{\rm cr} -1}
\right) \ll  \Lambda   \quad ( \alpha_\ast \simeq \alpha_{\rm cr}) \, ,
\label{Appel}
\end{equation}
which is based on  the same equation as the ladder SD  equation  with $\alpha(Q) \simeq \alpha_*$ and hence the
same form as the Miransky scaling \cite{Miransky:1984ef}, Eq.(\ref{Miransky}),  
of the Maskawa-Nakajima solution \cite{Fukuda:1976zb,Miransky:1984ef} .  We also have the same result as Eqs. (\ref{asymp},\ref{gammam}): %\cite{Yamawaki:1985zg} :
\begin{eqnarray}
\Sigma(Q) \sim 1/Q \,,\quad 
\gamma_m = 1 \quad {\rm at} \,\,\,\,\alpha_*=\alpha_{\rm cr}\, .
\end{eqnarray}

\begin{wrapfigure}{r}{6.6cm}
  \includegraphics[height=3cm]{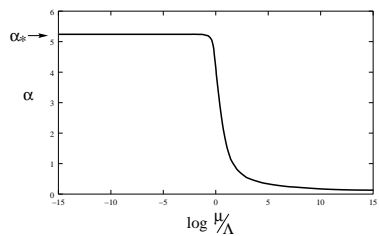}
\caption{Two-loop
running coupling of the large $N_f$ QCD ($N_f = 9, N_c =3$). At $Q \simeq m \ll \Lambda$, the coupling grows quickly
(not visible in this figure).}
\label{fig:step}
\end{wrapfigure}
Note that the exact BZ IR fixed point no longer exists in the SSB phase where the fermions acquire mass
and hence get decoupled  from the beta function, 
namely the gauge coupling quickly runs/grows up (strong asymptotically-free) for $Q<%\Sigma(0)
m \, (\ll \Lambda)$. Nevertheless, remnant of  the IR fixed point,  $\alpha (Q) \simeq \alpha_\ast$, dictates the coupling to
walk for a wide region
$%\Sigma(0) 
m < Q<\Lambda$ which is the  region most relevant to the physics of TC. See Fig.~\ref{fig:step}.

Such a large separation between the S$\chi$SB scale and the intrinsic scale of
the theory, $%\Sigma (0) 
m \ll \Lambda$,  is a salient feature
of the walking/conformal gauge theory in contrast  with the situation in 
the ordinary QCD where $%\Sigma(0) 
m \sim \Lambda_{\rm QCD}$.
Then the scale $\Lambda$
(although an analogue of $\Lambda_{\rm QCD}$) plays a role of cutoff in the SD equation and hence we may set the situation  $\Lambda \simeq \Lambda_{\rm ETC}$, while $m%\Sigma (0)
$ is directly tied to the weak scale $F_\pi =246 \, {\rm GeV}$ and hence plays a role of $\Lambda_{\rm TC}  (\ll \Lambda_{\rm ETC})$ in the previous discussions:  $m \rightarrow \Lambda_{\rm TC}$.

\subsection{Conformal Phase Transition}

Such an {\it essential singularity}  scaling law  like Eq.(\ref{Miransky},\ref{Appel}) characterizes an unusual phase transition, what we called ``{\it Conformal Phase Transition (CPT)}'',
where the Ginzburg-Landau effective theory breaks down \cite{Miransky:1996pd}: Although it is a second order (continuous) phase transition where the order parameter 
$m$ ($\alpha_*> \alpha_{\rm cr}$) is continuously changed to $m=0$ in the symmetric phase (conformal window, $\alpha_* < \alpha_{\rm cr}$), the spectra do not, i.e., while there exist light composite particles whose mass  vanishes at the critical point when approached from the side of  the SSB phase, 
no isolated light particles do not exist \cite{Banks:nn} in the conformal window, recently dubbed as ``unparticle''\cite{Georgi:2007ek}. 
 This reflects the feature of the conformal symmetry in the conformal window. 
In fact explicit computations show 
no light (composite) spectra in the conformal window,  in sharp contrast to the SSB phase
where light composite spectra do exist with vanishing mass as we approach the conformal
window $N_f \nearrow N_f^{\rm cr} $ \cite{Appelquist:1996dq,Miransky:1996pd, Harada:2003dc}.  

The essence of CPT may be illustrated by a simpler model, 
2-dimensional Gross-Neveu Model.
This is the $D\rightarrow 2$ limit of the $D$-dimensional Gross-Neveu model  ($2<D<4$) which has 
 the beta function and the anomalous dimension:~\cite{Kikukawa:1989fw, Kondo:1992sq}
\begin{eqnarray}
\beta(g) =-2g (g-g_*),\quad
\gamma_m = 2 g \, ,
\label{betagammaDNJL}
\end{eqnarray}
where  $g=g_*(\equiv D/2-1)=g_{\rm cr}$ and $g=0$ are respectively the UV and IR fixed points of the 
dimensionless four-fermion coupling, $g$, properly normalized (as  $g_* =1$ for  the $D=4$ NJL model).
There exist light composites $\pi, \sigma$ near the UV fixed point $g\simeq g_*$ in both sides of symmetric ($0<g<g_*$) and SSB 
($g>g_*$)  phases as in the NJL model. 

Now we consider   $D\rightarrow 2$ ( $g_* \rightarrow 0$) where 
we have a well-known effective potential: 
$V(\sigma , \pi) \sim (1/g-1)  \rho^2 +\rho^2 
 \ln (\rho^2/\Lambda^2) $,
or  $\partial^2 V/\partial \rho^2|_{\rho=0} = -\infty$, where $\rho^2=\pi^2+\sigma^2$. This implies breakdown of the Ginzburg-Landau theory which 
 distinguishes the SSB ($<0$) and  symmetric ($>0$) phases by the signature of the finite $\partial^2 V/\partial \rho^2 $ at the critical point $g=0$. 
Eq. (\ref{betagammaDNJL} ) now reads:
\begin{eqnarray}
\beta(g) =-2 g^2\, ,\, \,\,\gamma_m|_{g=0}= 0  \quad \quad (D= 2)\,,
\end{eqnarray}
namely  a fusion of the
UV and IR fixed points at $g=0$.
Now the symmetric phase is squeezed out to the region  $g<0$ (conformal phase) which corresponds to a {\it repulsive} four-fermion interaction and no composite states exist, while in the SSB phase ($g>0$) there exists a composite state $\sigma$ of  mass $M =2 m$ where $m$ is the dynamical mass
of the fermion 
$m^2 \sim \Lambda^2 \exp(-1/g) \rightarrow 0 \,\, (g\rightarrow + 0)$, which shows an essential singularity scaling, in accord with the beta function $\beta (g) = \Lambda \partial g/\partial \Lambda=-2 g^2$. Note the would-be
composite mass in the symmetric phase $|M|^2 \sim  \Lambda^2 \exp(-1/g)\rightarrow \infty \,\, (g\rightarrow -0)$.

Now look at the walking/conformal TC as modeled by the large $N_f$ QCD: When the walking coupling  is close to the critical coupling, $\alpha(Q) \simeq \alpha_* =\alpha_{\rm cr} $,
we should include the {\it induced}  four-fermion interaction which becomes relevant operator due to the anomalous dimension  $\gamma_m =1$, and  the system becomes 
``gauged Nambu-Jona-Lasinio'' model \cite{Bardeen:1985sm} whose solution in the full parameter space
was obtained in Ref.\cite{Kondo:1988qd}  

Thus we may regard the  {\it Walking/Conformal TC as the gauged Nambu-Jona-Lasinio model}.
Based on the solution\cite{Kondo:1988qd}, the RGE flow in $(\alpha,g)$ space was found to be along the  line of  $\alpha =\alpha_*$
($\alpha$ does not run),  on which  the (properly normalized dimensionless) four-fermion coupling $g$
runs, with the beta function and anomalous dimension given by ~\cite{Kondo:1991yk} \cite{Kondo:1992sq,Aoki:1999dv}
 \begin{eqnarray}
\beta(g) =-2(g-g_{(+)})(g-g_{(-)}), \quad 
\gamma_m = 2g +
\alpha_*/(2\alpha_{\rm cr})\, ,
\label{betagammaGNJL}
\end{eqnarray}
where    $g=g_{(\pm)}\equiv 
(1\pm \sqrt{1-\alpha_*/\alpha_{\rm cr}})^2/4$ are the UV/IR fixed 
points (fixed lines) for $\alpha_*\le \alpha_{\rm cr}$. The anomalous dimension takes the value $\gamma_m= 1\pm \sqrt{1-\alpha_*/\alpha_{\rm cr}} $ at the UV/IR fixed lines. Light composite spectra only exist near the UV
fixed line $g\simeq g_{(+)}$ in both SSB ($g>g_{(+)}$) and symmetric ($g>g_{(+)}$) phases as in NJL model.

Thus it follows  that as $\alpha_*\rightarrow \alpha_{\rm cr}$ Eq. (\ref{betagammaGNJL})  takes the form
\begin{eqnarray}
\beta(g) = -2 (g-g_*)^2 \,, \,\,\,\gamma_m |_{g=g_*}= 1\, , \quad\quad ( \alpha_* = \alpha_{\rm cr} ) \, ,
\end{eqnarray}
with $g_{(\pm)}\rightarrow 1/4 \equiv g_*$, and hence we again got a fusion of UV and IR fixed lines with the essential-singularity scaling
of  $m^2 \sim \Lambda^2 \exp (-1/(g-g_*)) $\cite{Kondo:1988qd} . Again there is a composite state with $M^2 \sim m^2  \rightarrow 0$
as $g\rightarrow g_* + 0$, while there are no composites    $|M|^2 \sim \Lambda^2 \exp (-1/(g-g_*))\rightarrow \infty$ for $g\rightarrow g_* - 0$.

The absence of the composites in the symmetric phase $g<g_*$ may be understood as in the 2-dimensional Gross-Neveu model for $g<0$,
namely the {\it repulsive} four-fermion interactions:
From the 
analysis of the RG flow, it was
 argued \cite{Kondo:1991yk}  
that the IR fixed line
$g= g_{(-)}$ is  due to the {\it induced} four-fermion interaction by the walking TC dynamics itself, while deviation from that line, $g- g_{(-)}$,  is  due to  
the  {\it additional} four-fermion interactions, repulsive 
($g<g_{(-)}$) and  attractive  ($g>g_{(-)}$), from UV dynamics other than the TC  (i.e., ETC). It is clear that no light composites exist for repulsive four-fermion interaction $g<g_{(-)}$, which becomes $g<g_*$ at $\alpha_*=\alpha_{\rm cr}$. 
 
\subsection{Top Quark Mass in the Walking/Conformal TC}
Note that  the anomalous dimension at the UV fixed line $g=g_{(+)} $ is even larger than
unity $\gamma_m  >1$ \cite{Miransky:1988gk},
due to the additional four-fermion coupling from other than the walking/conformal gauge dynamics, i.e., ETC. 
When $g(=g_{\rm ETC} +
g_{(-)}  )
> g_{(+)}$, or $g_{\rm ETC} >g_{\rm ETC}^{\rm cr}\equiv  g_{(+)}- g_{(-)}=\sqrt{1-\alpha_*/\alpha_{\rm cr}}$, SSB takes place with
large anomalous dimension  $\gamma_m =   1+ \sqrt{1-\alpha_*/\alpha_{\rm cr}} >1$, which would enhance the condensate more dramatically
and accommodate  the large top quark mass within the TC framework. 

If, on the other hand, the top quark instead of techni-fermion has a strong four-fermion interaction with scale $\Lambda_t$ close to the critical line but subcritical, $g=g_{(+)} -\epsilon$, 
then the ETC-induced top mass $m_t^{({\rm ETC})} (\ll m_t)$ will be enhanced by the anomalous dimension $\gamma_m   \simeq 2$, or $Z_m^{-1}=(\Lambda_{t} /m_t)^2$ (See Eq. (\ref{betagammaGNJL}). Thus we have  the realistic top mass $m_t =Z_m^{-1}\cdot m_t^{({\rm ETC})}$  without producing the light top-pion (See Subsection \ref{TMWCTC}).\cite{Mendel:1991cx}

\subsection{Confronting $S,T,U$  Parameters}
\label{Confronting}
Now the next problem is the so-called $S,T,U$ parameters \cite{PeskinTakeuchi} measuring possible new physics in terms of  the deviation of the LEP precision experiments  from the SM. In particular, $S$ parameter 
excludes the TC as a simple scale-up of QCD which yields $S = (N_f/2)  \cdot \hat{S}$ with
$\hat{S}_{\rm QCD} =0.33 \pm 0.04$. For a typical ETC model with one-family TC, $N_f=8$, %\cite{techni}
 we would get $S =
{\cal O} (1)$ which is  much larger than the experiments $S < 0.1$. This is the reason why many people  believe that the TC is dead. However, since the simple scale-up of QCD was already ruled out by the FCNC
as was discussed before, the real problem is whether or not the walking/conformal TC which 
solved the FCNC problem is also consistent with the S parameter constraint above.  

There have been many arguments \cite{Sundrum:1991rf} that the $S$ parameter value could be reduced in the walking/conformal TC than in the simple scale-up of  QCD. 

\begin{figure}[h]
   \begin{center}
     \includegraphics[height=2cm]{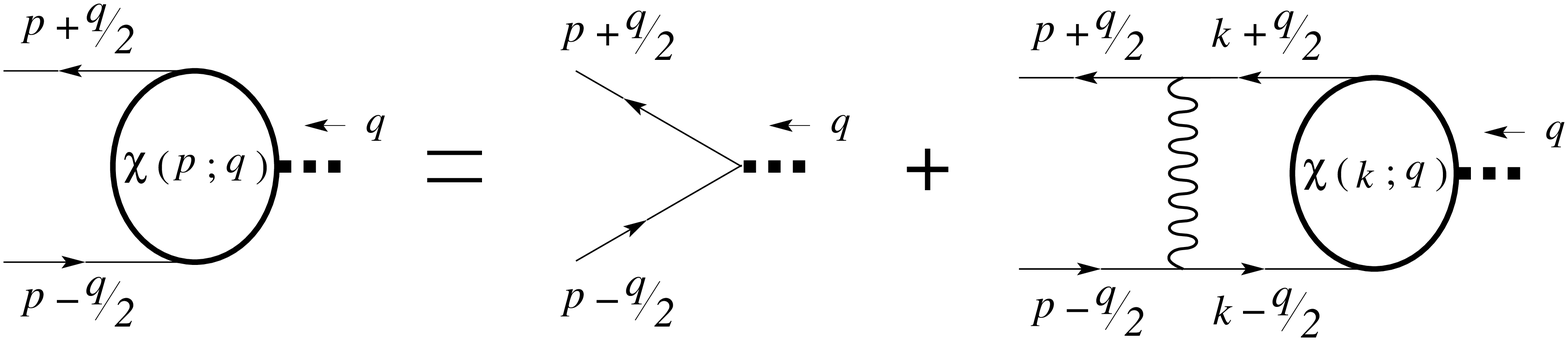}
   \end{center}
\caption{Graphical expression of the BS equation in the 
ladder approximation.}
\label{fig:IBSeq}
 \end{figure}

Here we present the most straightforward computation of the $S$ parameter for the large $N_f$ QCD
(for $N_c =3$) ,  based on the SD equation and (inhomogeneous) 
BS equation in the ladder approximation. \cite{Harada:2005ru} (see Fig.\,\ref{fig:IBSeq}). 
The $S$ parameter is given by $\hat S \equiv S/(2N_f) =-4\pi [\frac{d}{dQ^2} (\Pi_{VV}-\Pi_{AA}) ]\Big|_{Q^2=0} $,
where $\Pi_{VV} (\Pi_{AA})$ is the vector (axialvector) current correlator, which is 
obtained by closing the fermion legs (solution of SD equation) of the BS amplitudes (solution of BS equation).
Although the region studied was only $0.89 <\alpha_*<1$,  still somewhat away from the critical point  $\alpha_*=\alpha_{\rm cr}= \pi/4=0.79$, the result shows that  $\hat S$ gradually decreases
$\hat S/N_c \simeq 0.30/N_c$ ($\alpha_*=1$) to $\hat S/N_c \simeq 0.25/N_c$  ($\alpha_*=0.89$)
 as we approach  the conformal window $\alpha_\ast \searrow \alpha_{\rm cr}$ 
($N_f \nearrow N_f^{\rm cr}$) and  is definitely smaller values   than that in the ordinary QCD.  
The reduction does not seem to be so dramatic so far, due to technical limitation for  the present computation to get further close to  the conformal window.  It is highly desirable to extend the computation further close to the conformal window. 

However,  the results may imply nontrivial: The ladder SD and BS method tends to overestimate $\hat S$ in QCD,
which could be understood as scale ambiguity of  $\Lambda_{\rm QCD}$,   $\Lambda_{\rm QCD} \simeq 725 {\rm MeV}$ to reproduce the realistic  value of QCD $\hat S \simeq 0.33$, while $\Lambda_{\rm QCD} \simeq 500 {\rm MeV}$ reproducing other quantities yields  $\hat S\simeq 0.47$. Thus the reduction can read ${\hat S}_{\rm QCD}^{\rm ladder} = 0.47 {\rm \to} {\hat S}_{{\rm large} N_f  }^{\rm ladder}= 0.25$ more than $40\%$ reduction!
Then the actual value near the conformal phase transition 
point should be properly re-scaled by a factor roughly   2/3 to compare the QCD value within the ladder SD/BS
equation:  If this is done, then the value could be 
\begin{eqnarray} 
  {\hat S}^{\rm (re-scaled)}/N_{\rm TC} \simeq 0.067 (\alpha_*=1.0) \to 0.056(\alpha_*=0.89)\,\,,
  \label{re-scaled}
\end{eqnarray} 
which could be barely  consistent with the experiments if $N_{\rm TC}=2$. Since the walking/conformal theories are strong coupling theories and the ladder approximation would be no more than a qualitative hint,  more reliable calculations 
  are certainly needed, including the lattice simulations,  before drawing a definite conclusion about the physics predictions.  We will see.

\subsection{Walking/Conformal Signatures in LHC?}

 Walking/conformal TC would predict several characteristic phenomena in TeV region to be tested in the ongoing Tevatron and the upcoming LHC. There are huge varieties of  the TC models on the market, which predict rather diverse phenomena. To be definite, however, here we shall take a concrete model of ETC \cite{Hill:2002ap}, one 
family model of Farhi and Susskind embedded into a typical $SU(N_{\rm TC}+3)$  ETC which 
consists of  $SU(N_{\rm TC})$ TC of the $N_f=8$ (one-family) techni-quarks/leptons and the $SU(3)$
horizontal  gauge group for 3 families of quarks and leptons. 

 Since the weak scale is $(246\, {\rm GeV})^2
= (N_f/2) \times F^2_\pi$ we have  $F_\pi = 123 \, {\rm GeV}$, half of the naive scale-up of the $N_f=2$ QCD. As already noted in the footnote, the estimate of $N_f^{\rm cr}$ in
the large $N_f$ QCD has some ambiguity,  $N_f^{\rm cr} \sim [6 - 12]\, (N_c/3)$ in the literature,
we may assume $N_f^{\rm cr} = 2 N_c - 4 N_c$, i.e.,  the walking/conformal TC, $N_f \simeq N_f^{\rm cr}$, for the one-family model with $N_f=8$ would be realized for  $2 \le  N_{\rm TC} \le 4$, or
$N_{\rm TC} = 3 \pm 1$.

Then the light spectra of the one-family TC
would be as follows:

\begin{itemize}
\item  Pseudo NG Bosons (Techni-pions)\\
 This is heavily model-dependent  (Some models have no such objects).   In the one-family model the SSB of the global chiral symmetry $SU(8)_{\rm L}
  \times    SU(8)_{\rm R}$ for $N_f=8$ flavors produces 63 NG bosons: 3 would-be NG bosons absorbed into W/Z bosons and 60 pseudo NG 
bosons acquiring mass from explicit breaking due to various gauge interactions other than the TC.   Most problematic pseudo NG bosons would be colorless ones,  ``techni-axions'',  which 
          may have mass mostly from ETC or Pati-Salam type interaction (if any). It takes the form
          $m^2_{\rm pNG} \sim \langle \bar F F \rangle^2_{\Lambda_{\rm ETC}}/(F^2_\pi \, \Lambda^2_{\rm ETC})
 %which 
%would be $
\sim (\Lambda_{\rm ETC}/\Lambda_{\rm TC})^{2\gamma_m} \cdot \Lambda^4_{\rm TC}/\Lambda^2_{\rm ETC} $
%\sim  ({\rm GeV})^2 $ in a simple QCD scale-up but was    
 which reads $ m^2_{\rm pNG} \sim \Lambda^2_{\rm TC} \sim (350 \, {\rm GeV})^2$ for
$\gamma_m =1$ \cite{Yamawaki:1985zg}.
\item  Vector/Axialvector Mesons (Techni-$\rho$/Techni-$a_1$)\\
   Since in the vicinity of the conformal window in the large $N_f$ QCD,
 only the low energy scale is the tiny dynamical mass
   of fermion $m  (\ll \Lambda)$, one would expect that mass of  techni-$\rho$/techni-$a_1$ would also
  be vanishingly small $m_{{\rm techni}-\rho/a_1} /\Lambda \sim m/\Lambda \searrow 0$ as 
  $N_f \nearrow N_f^{\rm cr}$. \cite{Chivukula:1996kg}
A straightforward calculation based on the
SD and (homogeneous) BS equation (without the first diagram in the right-hand side of  Fig. \ref{fig:IBSeq}) in fact yields $m_{{\rm techni}-\rho}/ F_\pi  \sim 11 $,
$m_{{\rm techni}-a_1}/F_\pi \sim 12 $, with  $F_\pi/\Lambda \sim m/\Lambda \rightarrow  0$
as $N_f \nearrow N_f^{\rm cr}$.\cite{Harada:2003dc}  The  ratio for  techni-$\rho$ appears somewhat larger than  that in the case of the $\rho$ meson in QCD $m_\rho/f_{\pi} \simeq 8.5$.  Hence in the one-family model with $N_{\rm TC} = 3 \pm 1$, $F_\pi \simeq 125\, {\rm GeV}$,
this would imply that  $m_{{\rm techni}-\rho/a_1} \simeq (11-12)\, F_\pi \simeq 1.3-1.5\, {\rm TeV}$. 
    
\item  Techni-dilaton  (Techni-sigma)\\
The original walking/conformal TC predicted  a (massive) dilaton, ``techni-dilaton'',  as a pseudo NG boson of the spontaneous breakdown of the approximate scale invariance. \cite{Yamawaki:1985zg}  This looks like a Higgs boson in the SM.  In the vicinity of
the conformal window of the large $N_f$ QCD, we also expect a massive dilaton\cite{Miransky:1996pd}, whose mass 
may be estimated in the gauged NJL model as $m_{\rm technidilaton} \simeq \sqrt{2} m$ \cite{Shuto:1989te}, which is
consistent with the 
straightforward ladder SD/BS computation \cite{Harada:2003dc} in the large $N_f$ QCD: The scalar mass sharply drops when approaching 
the conformal window,  $m_{\rm technidilaton}\searrow  4 F_\pi
\simeq 1.5 m \simeq \sqrt{2} m$ as $N_f \nearrow N_f^{\rm cr}$. This  would imply that $m_{\rm technidilaton}
\simeq 500\, {\rm GeV}$ in the case of the one-family TC model. 
\end{itemize}

It should be emphasized that the dynamics near conformal window of large $N_f$ QCD
is strong coupling and hence the value of the BZ  IR fixed point and 
the critical coupling in the ladder approximation would be no more than a qualitative
hint. Before drawing a definite phenomenological  conclusion, 
we would need more reliable calculations of  the existence of the BZ IR fixed point, $N_f^{\rm cr}$ and the $S$ parameter and spectra such as in
 the lattice calculations.

\section{Hidden Local Symmetry (HLS) and Holography in the  Walking/Conformal Theories}

 Since at this moment there are some limitations on directly solving the strong coupling gauge 
theories near conformal window, we may take a different approach, namely, an effective
field theory. In contrast to the underlying microscopic theory, the effective filed theory consists
of  quantum fields for the light composite spectra as the dynamical variables.
 Here we
take the Hidden Local Symmetry (HLS) model \cite{Bando:1984ej} which extends the Callan-Coleman-Wess-Zumino (CCWZ) construction of the nonlinear sigma
model consisting of $\pi$ so as to incorporate  $\rho, a_1, ...$ as {\it composite gauge bosons}.   
 Note that the HLS is the induced gauge symmetry at the composite level which 
does not exist in the underlying theory. There is nothing wrong with this, since the gauge symmetry is not a symmetry.
The concept of HLS is often described by a later notion, Moose \cite{Georgi:1985hf}
 (actually the condensed Moose \cite{Arkani-Hamed:2001ca}).

\subsection{Hidden Local Symmetry}
It is generally shown \cite{Bando:1984ej}
that the nonlinear sigma model based on the coset space $G/H$
 is gauge equivalent to another 
model having a symmetry
 $G_{\rm global} \times H_{\rm local}$,
where the gauge symmetry
 $H_{\rm local}$ (``Hidden Local Symmetry'')  as well as the global symmetry $G_{\rm global} $ 
is spontaneously broken to give rise to a mass  $m_\rho$ of the gauge boson $\rho$
(``$\rho$ meson'' and the flavor partners),   and the left-over  (global) symmetry $H$ is a diagonal sum of the $H_{\rm global} (\in
G_{\rm global})$ and   $H_{\rm local}$, which is  identified with the original 
symmetry $H$ of the $G/H$. The latter model consists of  massless $\pi$ and massive $\rho$.

In the case of QCD with $N_f$ massless flavors, 
the CCWZ Lagrangian takes the form
${\cal L}_{\rm CCWZ}$ $= (F_\pi^2/4)\, {\rm tr} (\partial_\mu U^\dagger \, \partial^\mu U)$, where
$U(\pi(x)) =e^{2i \pi(x)/F_\pi}=\xi\, \xi$, with $\xi/\xi^\dagger \equiv e^{\pm i \pi(x)/F_\pi}$, which transform as $U(\pi)\rightarrow U^\prime= U(\pi^\prime) = g_L\, U(\pi)\, g_R^\dagger$, $(\xi^\dagger,\, \xi) \rightarrow ( \xi^{\dagger}, \, \xi)^\prime = h(g,\pi(x))\, (\xi^\dagger, \,\xi) \, g_{L,R}^\dagger$, with $g=(g_L, g_R) \in SU (N_f)_{L,R}$ and $h \in SU(N_f)_V$. 

Now, we may rewrite $U=\xi \, \xi$ into the form  $U =\xi_L^\dagger(x) \, \xi_R(x)$, where
$\xi_{L,R} (x) \rightarrow \xi_{L,R}^\prime(x)  = h(x) \, \xi_{L,R} (x) \, g_{L,R}^\dagger$, with $h(x) \in H_{\rm local}=
[SU (N_f)_{L+R}]_{\rm local},\,
g_{L,R} \in G_{\rm global} =[SU (N_f)_{L}\times SU (N_f)_{R}]_{\rm global}$. Here we have introduced a gauge symmetry  $H_{\rm local}$ as an ambiguity of dividing $U$ into two parts, which is independent of  the global symmetry $G_{\rm global}$. Thus a theory of $\xi_{L,R} (x)$ has a symmetry
$G_{\rm global} \times H_{\rm local}$ larger than that of the  original model, whose lowest order Lagrangian takes the form
\begin{eqnarray}
{\cal L}_{\rm HLS}&=&{\cal L}_{\rm A}+  a\, {\cal L}_{\rm V} - (1/2g^2)\, {\rm tr} F_{\mu\nu}^2,\\
\nonumber
{\cal L}_{{\rm A}/{\rm V}} &=&  -(F_\pi^2/4) \, {\rm tr} (D_\mu \xi_{\rm L}\cdot \xi_{\rm L}^\dagger \mp  D_\mu \xi_{\rm R}\cdot \xi_{\rm R}^\dagger)^2\, ,
\end{eqnarray}
where  $D_\mu \xi_{\rm L, R}=(\partial_\mu - i V_\mu)\xi_{\rm L,R}$, $a$ is a parameter and  $F_{\mu\nu}$ is the field strength of the HLS gauge field  $V_\mu$ of   $H_{\rm local}$ ($\rho$ meson)  which 
transforms as $V_\mu \rightarrow V_\mu^\prime = h(x)\, V_\mu \, h^\dagger (x) - i \partial_\mu \, h(x) \cdot
 h^\dagger (x)$, and $g$ is the HLS coupling constant. We can further gauge
(a part of) the $G_{\rm global}$ as $D_\mu \xi_{\rm L, R}=(\partial_\mu - i V_\mu)\xi_{\rm L,R} + i \xi_{\rm L,R} ({\cal L}_\mu, {\cal R}_\mu)$, where ${\cal L}_\mu, {\cal R}_\mu$ contain $\gamma$ and $W,\, Z$.
 
We now fix the gauge of   $H_{\rm local}$ as 
 $\xi^\dagger_L=\xi_R=\xi$ (unitary gauge) ,  so that  $(\xi_L, \, \xi_R)$ coincide with the CCWZ bases  $(\xi^\dagger,\,  \xi)$, and $g_{L,R}$ and $h(x) (= h(g,\pi(x))$ are no longer independent of each other,
leaving the symmetry $G$ spontaneously broken to $H$ which is a diagonal subgroup of $H_{\rm global}
(\subset G_{\rm global})$ and $H_{\rm local}$.  It is easy to see that ${\cal L}_{\rm A}$
is the same as the original nonlinear sigma model ${\cal L}_{\rm CCWZ}$ with $G$ gauged by
${\cal L}_\mu, {\cal R}_\mu$, while  $a\, {\cal L}_{\rm V} $
contains mass term of $V_\mu$ with a mass $m_\rho^2 = a g^2 F^2_\pi=g^2 F^2_\rho$, ``photon mass''
$m^2_\gamma= a e^2 F^2_\pi$,  $\rho-\gamma$ mass mixing $m^2_{\rho-\gamma}= e g_\rho, \,\, g_\rho=a g F^2_\pi $ and  $\rho-\pi-\pi$ coupling $g_{\rho\pi\pi}= (a/2) g$.  The direct $\gamma-\pi-\pi$ coupling comes from both  ${\cal L}_{\rm A}$ and   $a\, {\cal L}_{\rm V} $, yielding $g_{\gamma\pi\pi}
= (1-a/2) e$. 
Then we  have a successful KSRF(I) relation as an $a$-independent result: $g_\rho= 2 g_{\rho\pi\pi} F^2_\pi$ and nice phenomenological results are obtained for $a=2$: $m^2_\rho
= 2 g_\rho\pi\pi F^2_\pi$ (KSRF (II)),  $g_{\rho\pi\pi}= g$ (universality),  $g_{\gamma\pi\pi} =0$
(vector meson dominance).

In the case of $N_f=2$ the mass term indicates $SU(2)_{\rm local} \times [U(1)_{\tau_3}]_{\rm global}$,
with $[U(1)_{\tau_3}]_{\rm global} (\subset G_{\rm global})$ now gauged by a photon coupling, which is 
spontaneously broken down to the $U(1)_{\rm em}$, with the true (diagonalized) photon mass being precisely 0. This is precisely the same Higgs mechanism as in the Standard Model, $\rho^\pm$ corresponding to $W^\pm$ and $\rho^0$ to $Z^0$: $m_{\rho^0}^2/m_{\rho^{\pm}}^2= 1+e^2/g^2$. 

 \subsection{Vector Manifestation: Weakly Coupled Composite Gauge Boson Realized near the  Conformal Window}
Composite gauge bosons are usually regarded as strongly coupled, as we know about the $\rho$ meson,  which
is actually a conceptual barrier against model building for the composite W and Z bosons. Here I will discuss such a possibility  as ``Vector Manifestation (VM)'' \cite{Harada:2000kb%,Harada:2003jx
}  realized at the CPT. Such a dynamical possibility may be applied to the composite W/Z boson model.
The VM of chiral symmetry was also vigorously advanced in the chiral phase transition of the hot and dense QCD \cite{Brown:2006vn}. 
 
Let us discuss again the  large $N_f$  QCD, where we have seen that there exists the  Banks-Zaks IR fixed point $\alpha_\ast$ and a conformal window $N_f^{\rm cr} < N_f < 11 N_c/2$ ($0 <\alpha_\ast < \alpha_{\rm cr}$) .
If we regard the HLS model as an effective field theory for the underlying large $N_f$ QCD, then
we may expect that the chiral phase transition also takes place in the HLS model for
a certain large  $N_f$ corresponding to the conformal window.  

It was found \cite{Harada:2000kb} that this is indeed the case, once we match the HLS model (the simplest model $G_{\rm global}\times H_{\rm local}$ for $\pi$ and $\rho$ mesons) with the
underlying large $N_f$ QCD through OPE for current correlators at some scale $\Lambda$ where each theory gives a reasonable description. 
In the  HLS Lagrangian matched in  this way (the bare HLS theory defined at $\Lambda$) 
the bare $\pi$ ``decay constant'' $F_\pi(\Lambda)$ is given by  $F^2_\pi(\Lambda) \simeq N_c \times (\Lambda/4\pi)^2 \ne 0$ even when the chiral restoration takes place $\langle \bar q q \rangle \rightarrow 0$ in the QCD side and hence looks as if  in the broken phase at the scale $\Lambda$.
Nevertheless,  it receives quantum corrections of $\pi$ and $\rho$ loops  including quadratic divergence, $ - (N_f/2) (\Lambda/4\pi)^2$, \footnote{
$\pi$ loop alone gives a factor $N_f$ instead of  $N_f/2$. Thus the HLS is essential to having a sensible
result.
}
which may change the phase of the full quantum theory  into the symmetric phase
when we increase $N_f$: The true decay constant $F_\pi(0)$ as an order parameter ($\pi$ pole residue)  is given as  
\begin{equation}
F_\pi^2(0) =F_\pi^2(\Lambda) - (N_f/2) (\Lambda/4\pi)^2
\rightarrow 0
\end{equation}
 for $N_f \nearrow N_f ^{\rm cr} \simeq 2 N_c=6$ (more detailed analysis yields $N_f ^{\rm cr} \simeq 5$
),  which is compared with a lattice value \cite{lattice}  $6< N_f^{\rm cr} <7$ 
and the value $N_f^{\rm cr} \sim 4 N_c$ \cite{Appelquist:1996dq} given by the ladder SD equation and the two-loop BZ IR fixed point. 
Thus the HLS theory also has a chiral phase transition at large $N_f$.  

In the limit of  $\langle \bar q q \rangle \rightarrow 0$ in the QCD side,  
the above matching is possible only 
when $a(\Lambda)=1$ and $g(\Lambda)\rightarrow 0$  (``Vector Limit'') \cite{Georgi:1989gp} ,  so  that  $m^2_\rho/F^2_\rho= g^2 \rightarrow 0$ with  $F_\pi^2 = F^2_\pi(0) \rightarrow 0$  ($N_f \nearrow N_f^{\rm cr}$). 
Thus we encounter a new situation of the chiral symmetry restoration (Wigner phase) :
The $\rho$ meson becomes massless, with the longitudinal $\rho$ (NG boson field)  degenerate with $\pi$ as a chiral partner. We called it ``Vector Manifestation (VM)'' of Wigner phase of chiral symmetry.\cite{Harada:2000kb
} The HLS coupling now vanishes at the conformal window. 

\subsection{Summing up HLS Towers, or  Holography}

If we apply the VM to the TC instead of  the composite $W/Z$ model, the HLS gauge boson is the techni-$\rho$. The current correlator dominated by $\pi,\rho$ takes the form $(\Pi_{VV}-\Pi_{AA})/Q^2 =  F_\rho^2/(Q^2+M_\rho^2) - F_\pi^2/Q^2$,
which  yields  the $S$ parameter:
$\hat S = 4\pi (F_\rho/M_\rho)^2$ $=4\pi/g^2$, with $g$ being the HLS gauge coupling for techni-$\rho$.
Thus we would have $\hat S \rightarrow \infty$ as  $N_f \nearrow N_f^{\rm cr}$, which is in  opposite direction  to
the straightforward ladder/BS equation calculation mentioned in Sec.\ref{Confronting}. 
This is due to the infrared divergence of  $\rho$ which at VM would become massless as a chiral partner of 
massless $\pi$, while the  $\pi$ contribution  accidentally drops out in our definition of $\hat S$ which is not 
identical to the quantity measured by the LEP precision experiments. In the case of QCD
infrared divergence due to massless $\pi$ also takes place in $\hat S$ (or $L_{10}$) which is regularized by
the pion mass in reality.
It would be nice to find a proper definition of $\hat S$ to keep track of the chiral partner between $\pi$ and $\rho$
and to be compared with the LEP experiments.

Alternatively, we may consider a generalized HLS model \cite{Bando:1984ej} including the (techni-) $a_1$ as well as (techni-) $\rho$: 
$\hat{S} =4\pi  
(
(
F_\rho/M_\rho
)^2 
-(
F_{a_1}/M_{a_1})^2 
) 
= (4\pi/g^2)(
 1-
(
b/(b+c))^2  
)
%\, , 
%\label{Shat:VD}
$,
where $b,c$ are the parameters of this generalized HLS model to be running at loop level.
The one loop contribution of this model is more involved~\cite{Harada:2005br}, 
which may suggest  a possibility of a fixed point of the HLS parameters for giving a vanishing $\hat S 
\sim c/g^2 \to 0$ due to cancellation among $\rho$ and $a_1$ contributions at the chiral restoration point.   

Now, we are interested in summing up higher resonances in HLS model. 
We can easily extend the HLS to incorporate higher vector resonances.\cite{Bando:1984ej}
 In
the low energy region with the momentum
$p\ll m_\rho$ where
the $\rho$ kinetic term may be ignored, we can 
integrate out  the gauge boson $\rho$, or by use of the equation of motion for $\rho$, 
we get  $a\, {\cal L}_{\rm V} =0 $, and hence we are left with ${\cal L}_{\rm A}$ which is nothing but 
 the original  nonlinear sigma model on the coset space $G/H$. 
Similarly, the $G_{\rm global} \times H_{\rm local}$ HLS model is gauge equivalent to
another model having a symmetry $G_{\rm global} \times G_{\rm local}$, with the  gauge symmetry $G_{\rm local}$
 spontaneously broken
down to $H_{\rm local}$ giving mass to the axialvector meson  $a_1$, where we have introduced another nonlinear sigma model (Higgs field)  
to be absorbed into $a_1$.  In the energy region $m_\rho < p \ll
m_{a_1}$  we may integrate out  $a_1$ and get back to the $G_{\rm global} \times H_{\rm local}$ model. We can go further to $G_{\rm global} \times H_{\rm local}\times G_{\rm local} $ model to incorporate the vector
meson $\rho^\prime$ (see Phys. Rep. in Ref. \cite{Bando:1984ej}).   In this way we can proceed indefinitely to incorporate higher 
vector/axialvector mesons by introducing infinite set of nonlinear sigma models (Higgs fields): $G_{\rm global} \times G_{\rm local} \times G_{\rm local}$,   $G_{\rm global} \times H_{\rm local} \times G_{\rm local} \times G_{\rm local} \cdots$. 
In the case of QCD, this chain of the larger HLS models may be
summarized as $SU(N_f)_L \times SU(N_f)_R/SU(N_f)_V$ $\Rightarrow$ $[SU(N_f)_L \times SU(N_f)_R]_{\rm global} \times
[SU(N_f)_V]_{\rm local}$ $\Rightarrow$ $ [SU(N_f)_L \times SU(N_f)_R]_{\rm global} \times [SU(N_f)_L \times SU(N_f)_R]_{\rm local} $ $\Rightarrow$  $ [SU(N_f)_L \times SU(N_f)_R]_{\rm global} \times [SU(N_f)_V]_{\rm local}\times [SU(N_f)_L \times SU(N_f)_R]_{\rm local}  \cdots$. 

Now we consider a special parameter choice:
$a=1$ and  $g =0$ (vector limit) \cite{Georgi:1989gp}, in which case $\xi_L$ and $\xi_R$ get decoupled and hence we get two independent nonlinear sigma models
$[SU(N_f)_{L^\prime} \times SU(N_f)_{L}/SU(N_f)_{L^\prime+L}] \times [SU(N_f)_{R^\prime} \times SU(N_f)_{R}/SU(N_f)_{R^\prime+R}] $, where $\xi_L \rightarrow \xi_L^\prime =
g_{L^\prime} \xi_L g_{L}^\dagger$, and similarly for $\xi_R$ . Conversely, switching on the HLS coupling get
the global symmetry $G^\prime=SU(N_f)_{L^\prime}\times SU(N_f)_{R^\prime} $ down to $H_{\rm local} =  
[SU(N_f)_{L^\prime+R^\prime}]_{\rm local}$ and 
the model is reduced to the HLS model having $H_{\rm local} 
\times G_{\rm global}$ with $ G_{\rm global} = [SU(N_f)_{L}\times SU(N_f)_{R}]_{\rm global}$.
In terms of  the (condensed) Moose language \cite{Georgi:1985hf}, this situation is identical to the ``3-site  linear moose'' : $[SU(N_f)_L]_{\rm global} - SU(N_f)_{\rm local} - [SU(N_f)_R]_{\rm global}$,
with the site of HLS $SU(N_f)_{\rm local}$ in the middle being circled and with the $G_{\rm global} $ split to both ends, $[SU(N_f)_L]_{\rm global} $ and  $[SU(N_f)_R]_{\rm global}$, being
circled (when gauged with external gauge fields ${\cal L}_\mu, {\cal R}_\mu$)  or opened (when ungauged) and the link denoted by ``$-$'' being each nonlinear sigma model base. In this sense Moose is nothing but a reformulation of HLS. 

\begin{figure}[h]
\begin{center}
    \includegraphics[trim = 0 350 100 120 , scale=.5, clip]{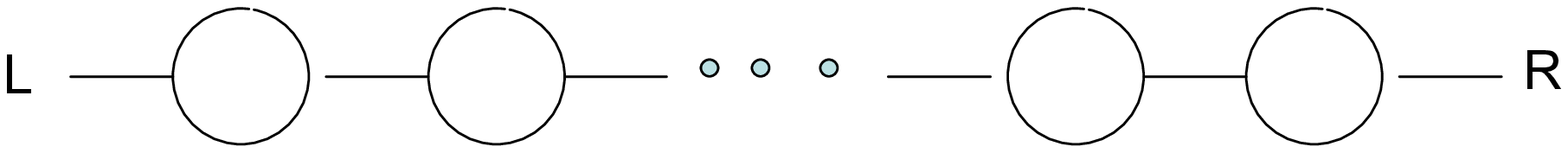}
  \end{center}
\caption{Linear moose for  arbitrary number of HLS's. Each circle stands for $SU(N_f)$ HLS, and 
each bar connecting two circles stands
for the  nonlinear sigma model transforming under the connecting two HLS's. Two end points stand for $[SU(N_f)_{L,R}]_{\rm global}$ which 
may be gauged by the external gauge fields ${\cal L}_\mu, {\cal R}_\mu$. .
}
\label{fig:moose}
\end{figure}

In a particular parameter choice (generalization of the choice of $a=1$ in the vector limit)  it is sometimes convenient to put it into the linear moose style (Fig.\ref{fig:moose}): 
$[SU(N_f)_L]_{\rm global} - [SU(N_f)_L]_{\rm local} - [SU(N_f)_L]_{\rm local} - \cdots - [SU(N_f)_R]_{\rm local}  - [SU(N_f)_R]_{\rm local}  - [SU(N_f)_R]_{\rm global}$.
By incorporating infinite tower of  massive HLS gauge bosons in this way, one
actually arrives at the  gauge theory in 5 dimensions \cite{Son:2003et}, 
the {\it massive} tower of HLS gauge bosons being nothing but the {\it Kaluza-Klein (KK) tower} of the {\it massless} 5-dimensional gauge boson $A^{M}(x^M)$ ($M=\mu, 5)$. 
When the 5-th dimension is deconstructed/latticized \cite{Arkani-Hamed:2001ca}, the Wilson line $e^{i \int dx_5 A^5}$ (link variable in the lattice gauge theory) acts like a
nonlinear sigma model (Higgs) corresponding to each ``$-$'' in the above moose
and is adsorbed into the gauge bosons  (KK modes) of the HLS at the nearest sites (circles). 
Actually, the holographic approach to QCD  gives
 the 5-dimensional gauge theory which is
nothing but an infinite set of HLS gauge bosons. \cite{Son:2003et,Sakai:2004cn}

Converse is true. We can always integrate out \cite{Harada:2006di}
the KK tower of the holographic model to get back to the
simplest HLS model with the lowest resonance, the $\rho$ meson, plus
 ${\cal O} (p^4)$ terms in the HLS chiral perturbation (See\cite{Harada:2000kb%,Harada:2003jx
} ) with definite coefficients.

\subsection{Holographic Walking/Conformal TC}
\begin{wrapfigure}{r}{6.5cm}%\begin{figure}
 % \begin{center}
  \includegraphics
[keepaspectratio=true,height=23mm]
{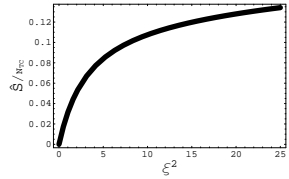}
  %\end{center} 
  \caption{Plot of $\xi^2$-dependence of $\hat{S}/N_{TC}$ with $\gamma_m \simeq 1$. 
The blob is the result of the ladder SD and BS equations, $\xi$ from homogeneous 
BS equation~\cite{Harada:2003dc} and $\hat S$ from inhomogeneous BS 
equation~\cite{Harada:2005ru} }
  \label{s-xi2-ad=1-ladder}
\end{wrapfigure} 
The reduction of $S$ parameter in the walking/conformal TC has  been argued  in  a version of  the holographic 
  QCD \cite{Erlich:2005qh}
deformed to the walking/conformal TC  by  tuning a parameter
to simulate the large anomalous dimension $\gamma_m \simeq 1$. \cite{Hong:2006si} 
We recently examined \cite{Haba:2008nz} such a possibility paying attention to the renormalization point dependence 
of the condensate. 
We explicitly calculated the $S$ parameter in entire parameter space 
of the holographic walking/conformal technicolor (See Fig. \ref{s-xi2-ad=1-ladder}) .

The  $S$ parameter was  given as a positive monotonic function 
of  $\xi$ which is fairly insensitive to $\gamma_m$ and 
continuously vanishes as $\hat S \sim \xi^2 \to 0$ when  $\xi \to 0$,   
where  $\xi$ is the (dimensionless) vacuum expectation value of the bulk scalar field at the infrared boundary of the 5th dimension 
$z=z_m$ and  is related to the mass of  (techni-) $\rho$ meson ($M_\rho$) and the decay constant ($f_\pi$) as 
$\xi \sim f_\pi z_m \sim f_\pi/M_\rho$ for $\xi \ll 1$. 
However, although $\xi$ is related to the techni-fermion condensate $\langle \bar F F \rangle$, 
we find no particular suppression of $\xi$ and hence of $S$ due to large $\gamma_m$,  
based on the correct identification of the renormalization-point dependence 
of  $\langle \bar F F \rangle$  in contrast to the literature\cite{Hong:2006si}. 

Curiously enough, a set of the values of $\xi^2$ (read from 
$F_\pi/M_\rho$\cite{Harada:2003dc}) and 
$\hat S/N_{{\rm TC}}$ in Eq.(\ref{re-scaled})~\cite{Harada:2005ru}  in the straightforward calculation of ladder SD/BS equations
fit in  the line of the holographic result in Fig.\ref{s-xi2-ad=1-ladder}.

\section{Top Quark Condensate in Walking/Conformal Theories}
Top Quark Condensate (Top Mode Standard Model, TMSM) \cite{Miransky:1988xi,Nambu89,BHL:1989ds}  is an idea alternative to the technicolor, and as such has potentiality to account for the origin of mass of all the SM particles.\cite{Miransky:1988xi}(For a recent attempt see \cite{Dobrescu:2008sz}).

The original top quark condensate model 
was formulated in the gauged Nambu-Jona-Lasinio (NJL) model, \cite{Miransky:1988xi}
 four-fermion theory plus Standard Model
gauge couplings, whose phase structure (critical line) was revealed in the ladder SD equation \cite{Kondo:1988qd}.
The gauged NJL model was shown \cite{Miransky:1988gk} to have a very large anomalous dimension $\gamma_m\simeq 2$ due to strong four-fermion interaction at the UV fixed line $g=g_{(+)}$ in Eq.(\ref{betagammaGNJL}) (identified with  the critical line) for small SM gauge coupling ($\alpha \sim 0$). 
The existence of critical line implies that a tiny difference among  the four-fermion 
coupling (or the gauge coupling)  of the top and bottom could result in $m_t \ne 0$ and
$m_b =0$, thus explaining a large hierarchy among top and bottom (and other quarks/leptons) :
$m_t \gg m_b, m_c \dots$. 
The model predicted (long before
the discovery of the top quark) $m_t \simeq 250 \, {\rm GeV}$ (large $N_c$ leading)  \cite{Miransky:1988xi} and
$m_t \simeq 220 \, {\rm GeV}$ (including $N_c$ subleading effects) \cite{BHL:1989ds} for  the cutoff
$\Lambda \sim 10^{16} - 10^{19}\, {\rm GeV}$. This reflects the reality at 0th order approximation that
only the top mass  (as well as $W,Z$ masses) is on the order of weak scale. However, if  $\Lambda$ is a natural
scale in TeV region, $m_t$ would be as large as $500  \, {\rm GeV}$.   
  The model predicts a Higgs boson as a bound state of $\bar t t $ (``Top-sigma'') \cite{Miransky:1988xi,Nambu89}
whose mass is $m_H \simeq 2 m_t$
in the NJL model, but is changed to
$m_H \simeq 1.1 m_t$
\cite{BHL:1989ds}  in the gauged NJL model.
Now the questions: What is the origin of the four-fermion interactions?  How can we get a realistic
top quark mass  $m_t \simeq 172 \,{\rm GeV}$ in a natural way?

\subsection{TMSM with extra dimensions}
Let us now come to an extension \cite{ACDH:2000hv,HTY:2000uk}
of  the TMSM as simply the 
SM formulated  in the higher dimensional bulk {\it without ad hoc four-fermion interactions}, where  the (dimensionless) color coupling in the bulk gets strong when the extra dimensions become operative in the high energies in TeV region. 
It was noted earlier \cite{Dobrescu:1998dg} that   the bulk gluon exchanges 
(or the gluon KK mode exchanges) play the role of   the four-fermion interactions triggering the
top quark condensate in the original model and the bulk top quark give rise to KK modes
of the top quark which  can bring the top mass prediction of the original model down to the realistic value $m_t \simeq 172 \,{\rm GeV}$ in a way similar to the top-seesaw\cite{Dobrescu:1997nm}. 

Actually, the QCD with compactified $(D-4)$ extra dimensions becomes a walking/conformal gauge 
theory having a {\it UV fixed point
with large anomalous dimension} $\gamma_m \simeq D/2-1$.\cite{HTY:2000uk}   
Because of the UV fixed point, the dimensionless bulk  QCD coupling does not grow indefinitely, while the $U(1)_Y$ bulk coupling has a Landau pole at high energy and hence dominates the QCD coupling at certain high energy scale to favor the $\tau$ condensate rather than the top quark condensate. Thus it is highly nontrivial whether
or not there exists a region where the top quark condensate is a Most Attractive Channel (MAC) favored to others.  We called such a region  a ``topped MAC (tMAC)'' region which is identified with
the cutoff where the composite Higgs of $\bar t t$ is formed. Actually the tMAC region
is so narrow, if existed at all, that we can predict the mass of  top quark and also of  Higgs boson as a  $\bar t t$ composite without much ambiguity: $m_t = 172-175 \, {\rm GeV}$ $m_H
= 176-188  \, {\rm GeV}$ ($D=8$ with  $R^{-1} =1-100 {\rm TeV}$
where $R\,  (=1/\Lambda)$ is  the radius of  compactified extra $(8-4=4)$ dimensions. \cite{HTY:2000uk}
If one assumes that SM particles live in the 6-dimensional brane (5-brane) in the $D=8$ bulk where
only the gluons live, one  gets: $m_t = 177-178 \, {\rm GeV}$ $m_H
= 183-207  \, {\rm GeV}$. The Higgs  in this characteristic mass range will  be discovered immediately once the LHC started.

\subsection{Top Mode Walking/Conformal TC}
\label{TMWCTC}
 The top quark is very special in the ETC, since the top and bottom mass given by  $m_{t/b} \sim \frac{1}{\Lambda_{\rm ETC}^2}\, \langle \bar F F \rangle_{\Lambda_{\rm ETC}}$ 
would  require large isospin violation in the condensate $\langle \bar F F \rangle_{\Lambda_{\rm ETC}}$
in order to produce large mass splitting $m_t \gg m_b$, which would conflict the $T$ parameter constraint.
A possible way out , so-called
``Topcolor-assisted TC (TC2)''\cite{Hill:1991at} is to introduce  the (isospin violating) top quark condensate, in addition to the 
(isospin conserving) TC condensate which gives main contribution to the weak scale (W/Z mass). Such a model generically predicts  (besides the top-sigma) a salient
 light pseudo NG boson, top-pion,
whose mass comes from the ETC-induced top/bottom mass. If we require the top mass mainly comes from
the top condensate rather than the ETC-origin, then we found \cite{Fukano:2008iv} that the mass of the top-pion $m_{\pi_t}$
is severely
constrained as $m_{\pi_t}< 70 \, {\rm GeV}$ due to the large anomalous dimension $\gamma_m \simeq 2$ 
of the dynamics for the top
quark condensate.

\section{Conclusion}
We have discussed  composite models for the electroweak 
sector, based on walking/conformal gauge theories with large anomalous dimension near
the IR/UV fixed point. Many such models predict Higgs-like composite spectra somewhat heavier than those 
anticipated in the typical  SUSY theories and SM, and hence will be distinguished  in the LHC experiments:
Walking/Conformal TC will have  techni-dilaton (techni-sigma) which is expected to have mass typically  around 
$\sqrt{2} M_f  (>  500-600 {\rm GeV}) $,  where $M_f$ is the techni-fermion mass. Top quark condensate will
have a top-sigma whose mass slightly less than twice top quark mass $< 2m_t \sim 350 {\rm GeV}$.
If LHC did not find light Higgs with mass lighter than, say $180 {\rm GeV}$, there will be a good chance for the composite
model and  history will repeat itself on the old avenue that Sakata, Nambu, and Maskawa  walked on. We will see.

 \section*{Acknowledgements}
We would like  to thank T. Kugo for hospitality at
Yukawa Institute for Theoretical Physics, Kyoto University.
Part of this work was done during  the 16th Yukawa International Seminar ``Particle Physics beyond the Standard Model'' held at Yukawa Institute of Theoretical Physics, Kyoto University, Jan. 26-March 25, 2009. 
This work was supported in part by the JSPS
Grant-in-Aid for Scientific Research (B) 18340059 and the Daiko Foundation,
 and by the Grant-in-Aid for Nagoya
University Global COE Program, "Quest for Fundamental Principles in the
Universe: from Particles to the Solar System and the Cosmos", from the Ministry
of Education, Culture, Sports, Science and Technology of Japan.

 \end{document}